\documentclass[conference, 10pt]{IEEEtran}
\IEEEoverridecommandlockouts

\usepackage{ifpdf}
\usepackage{tabularx}
\usepackage{threeparttable}
\usepackage{makecell}
\usepackage{xcolor, soul}
\usepackage{subcaption}
\usepackage{rotating}
\usepackage{enumitem}

\definecolor{gray}{gray}{0.8}
\newcommand{\boxfix}[2][]{%
  \ifthenelse%
    {\boolean{showcomments}}%
    {\colorbox{gray}{\parbox{0.95\linewidth}{\emph{\normalsize\sffamily{\bfseries\/Comment\ifthenelse{\equal{#1}{}}{:}{~#1:}}~#2}\/}}}%
    {}%
}

\usepackage{booktabs}
\usepackage{amsfonts}
\usepackage{hyperref} 
\usepackage{url}

\newcommand{\takeaways}[1]{
\noindent \fcolorbox{black}{white}{\begin{minipage}[c]{0.96\columnwidth}
\textbf{Takeaways}: #1
\end{minipage}}
}

\begin{document}

\title{A Comprehensive Quantification of \\ Inconsistencies in Memory Dumps}

\author{
Andrea Oliveri
\IEEEauthorblockA{ \\
\textit{EURECOM}\\
Biot, France \\
andrea.oliveri@eurecom.fr}

\and
Davide Balzarotti
\IEEEauthorblockA{ \\
\textit{EURECOM}\\
Biot, France \\
davide.balzarotti@eurecom.fr}
}


\maketitle

\begin{abstract}
Memory forensics is a powerful technique commonly adopted to investigate compromised machines and to detect stealthy computer attacks that do not store data on non-volatile storage. To employ this technique effectively, the analyst has to first acquire a faithful copy of the system's volatile memory after the incident. However, almost all memory acquisition tools capture the content of physical memory without stopping the system's activity and by following the ascending order of the physical pages, which can lead to inconsistencies and errors in the dump. In this paper we developed a system to track all write operations performed by the OS kernel during a memory acquisition process. This allows us to quantify, for the first time, the exact number and type of inconsistencies observed in memory dumps. We examine the runtime activity of three different operating systems and the way they manage physical memory. Then, focusing on Linux, we quantify how different acquisition modes, file systems, and hardware targets influence the frequency of kernel writes during the dump. We also analyze the impact of inconsistencies on the reconstruction of page tables and major kernel data structures used by Volatility to extract forensic artifacts. Our results show that inconsistencies are very common and that their presence can undermine the reliability and validity of memory forensics analysis.
\end{abstract}

\begin{IEEEkeywords}
memory forensics, data-structures, memory acquisition, inconsistencies
\end{IEEEkeywords}

\section{Introduction}
\label{sec:introduction}

Memory forensics is a powerful technique that can provide valuable insights into the behavior and state of a computer system. It is routinely used to investigate and respond to computer incidents and as part of threat hunting, malware analysis, and intrusion detection procedures. Most research efforts in this area have been dedicated to reconstructing an accurate representation of the content of the memory~\cite{dolangavitt,dimsum,lin2010automatic,gu2014multi,siggraph,urbina2014sigpath,feng2014mace,deepmem,mercier2017dynstruct,slowinska2010dde,troshina2010reconstruction}. However, before analyzing the memory an analyst has to first acquire it.

Several approaches can be adopted to acquire a copy of the physical memory of a bare-metal machine. One option is to rely on specialized hardware devices that can access the content of the RAM over DMA. Another popular option is to use software-based memory acquisition tools that run within the target system. In this case, the analyst installs specialized software or leverages an integrated functionality of the operating system to copy the contents of the system's RAM onto an external storage device or transmit it over the network. Over time, multiple solutions~\cite{reina2012hardware,latzo2021bringing,lime,avml} have been developed to accomplish this task. In most cases, they rely on dedicated kernel modules that acquire the memory without altering the kernel behavior and interrupting its execution.

Independently of the fact that the memory is acquired by a software module or a hardware device, the acquisition process usually retrieves the content of physical pages in ascending order, according to their address in the physical address space. This technique is easy to implement and minimizes the logic required to track which pages have already been acquired, thus reducing the memory footprint of the acquisition tool. For this reason, this is also the technique adopted by almost all currently existing acquisition software for major operating systems such as Windows and Linux, and it is the only technique that can be used when the OS is unknown. In fact, in the context of the OS-agnostic memory forensics~\cite{fossil}, the lack of information about the structure of the operating system prevents the analyst from giving precedence to those pages that contain information of potential interest for a forensic analysis~\cite{pagani2019introducing}, or that can be more easily altered.

A common misconception is that the content of a memory dump faithfully reflects the content of the memory at the moment of its capture. In reality, on a bare-metal machine\footnote{This is not true for virtual machines, where the hypervisor can suspend the guest execution during the acquisition process.} memory acquisition occurs while the system is still running. Consequently, memory dumps are not true atomic snapshots of the system's state but rather a collection of memory pages captured at different moments in time. This can lead to inconsistencies in the analyzed data and errors in the investigation results. 
For instance, researchers have reported inconsistencies in page tables in at least 20\% of the acquired images~\cite{case2017memory} and a corrupted image once every five acquisitions~\cite{stefan2018}. 
This problem was first studied from a theoretical perspective in 2012 by \textit{Vomel and Freiling}~\cite{vomel2012correctness}, who introduced a set of definitions to describe the various requirements that a memory image must meet in order to be considered a faithful copy of the system's memory at a given point in time. 
The same authors also tested various acquisition software and confirmed the presence of inconsistencies in the dumps they produced~\cite{gruhn2016evaluating, vomel2013evaluation}. 
More recently, \textit{Pagani et al.}~\cite{pagani2019introducing} investigated how to assess the atomicity of data structures commonly used by Volatility~\cite{volatility} to help the analyst during the forensics analysis. 
However, even though these types of inconsistencies have been discussed in numerous papers over the past fifteen years, 
their presence and assessment have always been anecdotal, and to date, there are no studies that precisely quantified the \emph{actual number} of inconsistencies in a memory dump and their impact on the results of a forensics analysis.

\noindent \textbf{Contribution} -- In this work, for the first time, we precisely quantify the number and type of inconsistencies observed in non-atomic memory dumps and pinpoint them to the exact field of the affected kernel data structure. We employ the PANDA~\cite{dolan2015repeatable} record-replay infrastructure to track all write operations the OS kernel performs during a memory acquisition process. Gathering comprehensive data over several experiments allows us to study how different acquisition techniques, target file systems, and operating systems influence the frequency of kernel writes. We proceed to classify and quantify potential types of inconsistencies that can impact page tables (page smearing) and evaluate the resulting implications for the accurate reconstruction of the virtual address spaces of the kernel and user-space processes. We finally track all the main kernel data structures that Volatility 2 uses to extract forensic artifacts from memory images. This allows us to assess how inconsistencies can undermine the reliability and validity of the results presented to the analyst.

Although the risk of inconsistencies was already well known within the forensic community, this paper is the first to provide a precise quantification of the problem, yielding disconcerting results. 
What the memory forensic community believed to be a sporadic inconvenience is instead a systemic problem that extensively affects \emph{all} memory dumps. 
In fact, our experiments show that the occasional errors analysts reported in their investigations were only the tip of the iceberg of a huge amount of inconsistencies that affect a significant fraction of pointers and kernel data structures. 
One of the pillars of digital forensics is that the analysis should be based on a faithful copy of the data. Our research shows that non-atomic acquisition methods violate this principle, not just by introducing minor differences but by providing a completely untrustworthy picture of the state of the memory. This quantification and systematization study aims to reveal the true scale of the problem and lay the groundwork for developing future acquisition techniques.

\section{Problem Statement}
\label{sec:approach}
A memory acquisition is considered \textit{atomic} if the content retrieved by a dumping tool, despite requiring a non-negligible amount of time, is indistinguishable from a hypothetical snapshot captured in a single, instantaneous operation.
 For practical purposes, as pointed out by \textit{Vomel and Freiling} in 2012~\cite{vomel2012correctness}, the definition can be relaxed to a snapshot of the memory which ``\textit{does not show any signs of concurrent system activity}''. This concept of atomicity allows for the collection of memory pages at different points in time as long as the acquisition procedure maintains the causal relationships between memory operations and inter-process synchronization primitives. Although this definition is remarkably elegant, it poses significant challenges regarding practical measurement as it is difficult to verify in practice. A more practical definition was introduced by \textit{Pagani et al.} in 2019~\cite{pagani2019introducing} and then formalized by \textit{Ottmann et al.}~\cite{freiling2021defining}. The authors said that a collection of physical pages, subject to some causal relations, is considered \textit{time-consistent} if there exists a hypothetical atomic acquisition process that could have yielded the same outcome. In other words, there was a specific moment during the acquisition process when the content of those pages coexisted in the system's memory. If we consider the set of all kernel structures and their relations as a graph, a time-consistent dump produces a graph locally equivalent to one obtained from an atomic dump. However, the two graphs differ at a global scale because the individual links between kernel structures retain their causality, but the pages containing the structures are not acquired all simultaneously. In order to produce a time-consistent dump, the imaging tool must be aware of the relationships that exist between the various pages of the physical memory and must be able to find a dump sequence that satisfies all of them. While this is possible in theory, it is not feasible in practice. Specifically, when tools dump pages in a sequential order (which is the case with most available tools for major operating systems), achieving time-consistent atomicity is impossible if the entire OS is not frozen during the acquisition process --- a condition that can only be met dumping a virtual machine from the hypervisor.

In a running system, there are two main causes of inconsistencies: the virtual-to-physical address translation and the internal and concurrent activity of the kernel and user space programs. In modern kernels that support virtual memory abstraction, each program operates within its own distinct and private address space known as the virtual address space. The virtual addresses used by a particular process are translated into physical memory locations where the data resides through the combined efforts of the Memory Management Unit (MMU) and the operating system. On Intel and ARM architectures, this translation process requires in-memory data structures, hierarchically organized, prepared by the kernel and used by the MMU: the page tables~\cite{mmushell}. Like all information contained in memory, page tables can also be affected by non-atomic acquisitions, a phenomenon called \textit{page table smearing}~\cite{case2017memory}. Page table smearing occurs when an acquired page table references page tables of lower levels whose content is modified by the kernel before they get acquired, resulting in inconsistencies and errors in the virtual-to-physical address translation, which is later performed by the forensic analysis tool.
As a result, the post-mortem analysis might miss entire regions of the virtual space, it can show inconsistent permissions bits (such as write permission, execution permission, or accessibility in the kernel or user space), or it can make errors in the reconstruction of the virtual memory of a process (e.g., by including data pages that originally belonged to other processes). This phenomenon can have serious repercussions on the results of the forensic analysis. Page table smearing can also occur in page tables that belong to kernel memory or areas shared among multiple processes (e.g., for pages that host dynamically linked libraries), further exacerbating the problem.

The second source of inconsistencies is the concurrent activity of the kernel and user processes during the memory acquisition process. The scheduling routine allocates the available CPU time among processes ensuring that each process is executed without monopolizing the system's resources. The kernel itself and all the other privileged software are subjected to these rules. So, when the analyst runs an acquisition tool, regardless of its level of privileges, its code runs along with other processes of the system\footnote{{If the CPU architecture supports more privileged modes than kernel mode, such as VMM and SMM modes in Intel CPUs, the execution of a acquisition tool in those modes cannot be interrupted by the kernel or user space programs, resulting in an atomic view of the memory used by lower privileged applications.}}. This means that the acquisition software cannot, in general, pause all other processes, including the kernel, to prevent them from modifying the memory content during the dump procedure. This can introduce numerous inconsistencies within a memory dump because data structures allocated in distant regions in the physical address space are dumped at different times due to the scheduling policy. Things are even worse in multicore systems, in which multiple processes can run simultaneously on different physical CPUs, further increasing the chances of inconsistency in the dump.

As a result, the kernel, with its ability to write to any physical page in the system and the ability to interrupt any process, is thus the major culprit in generating inconsistencies within a memory dump. Moreover, since memory forensic analysis of a system always requires the analysis of kernel data structures, in this work, we will focus only on inconsistencies in the address space of the kernel itself.

\section{Measurement Technique}
\label{sec:measure_technique}
In our work, we want to quantify the inconsistencies that can be introduced by the kernel and the acquisition tool in the content and links among different \textit{structures}. Structures are blocks of bytes that store data and have causal relations with other structures in memory. To clarify, they can be imagined as C \texttt{struct}s containing data and pointers to other structures of the same or different type. It is important to note that this definition includes both data structures used by the kernel linked by pointers containing virtual addresses and also page tables used by the operating system and the MMU to translate virtual addresses into physical ones. In the latter case, the structures are the page tables themselves, and their entries can be seen as pointers in physical address space to other lower-level tables.

We consider the acquisition of a single 4KiB page as an atomic event because all the acquisition tools treat pages as atomic units and use a single call to the kernel to copy their content. In fact, kernel APIs within the low-level memory management system allow for mapping, allocating, and freeing only entire pages, which is a direct consequence of the functioning of the MMU.

To collect information about possible inconsistencies, it is essential to observe and monitor the kernel's activity throughout the entire dumping process.
Since this cannot be done directly on a bare-metal machine without using specialized hardware~\cite{latzo2021leveraging}, in our approach, we use PANDA~\cite{dolan2015repeatable}, an open-source tool based on QEMU that offers the ability to record and replay executions of an entire virtual machine. During
the replay of an execution trace, PANDA allows, through its plugin framework, to perform sophisticated analysis and hook the system in many different ways
(e.g., by triggering a callback when the CPU executes instructions at certain addresses or by intercepting single writes on memory and reads the content of CPU
registers).

In our experiments, we record the execution of an operating system under various conditions, such as during a memory acquisition process using a kernel module, as if it were performed on a bare-metal machine. Then, we replay the execution traces and analyze them with a set of custom plugins we designed to collect statistics about the activity of the kernel, the acquisition tool, and the state of data structures that we want to track. In particular, to monitor the evolution of the relations among structures during memory acquisition, we implemented a technique based on a versioning system. Before starting the replay we analyze the virtual machine memory to identify all the structures we want to track and explore their relations. For each tracked structure that appears in memory, we save the timestamp of its allocation and deallocation, and we assign it a unique identifier. During the replay of the VM execution, whenever there is a modification to an interesting field or a pointer within a structure, we record the operation and, if necessary, explore the newly pointed structures. In addition, for each pointer that we track, we also maintain the unique index of the structure it points to. We use a unique index instead of its address because the kernel can allocate and deallocate different objects at the same memory location. When a page is dumped, we freeze the state of all the structures it contains: for each structure, we save the value of its fields and the value of the fields of the pointed structures. In this way, at the end of the replay, we can detect inconsistencies and also discriminate among their types by checking whether the content of a pointed structure, saved when the page that contained it was acquired, is equal to its content saved when the pointing structure was acquired.

To study inconsistencies in memory dumps, we follow a bottom-up approach, starting from individual pointers and moving up to the composite kernel data structures that are used in forensics analysis. We start, in Section~\ref{res:idle_system}, measuring the activity of the kernels of three different OSs in idle state. Then, in Section~\ref{res:other_oss}, simulating a process of dump, we collect statistics for the same three OSs about the relations among physical pages containing kernel pointers and how their physical page allocation strategy influences the number of inconsistencies in the memory image. In Section~\ref{sec:technique}, we study how the tool, the different acquisition modes, and the target filesystem influence the content of the memory snapshot on Linux systems. Then, in Section~\ref{sec:approach_inconsistencies}, we classify the different types of inconsistencies involving kernel structures that can appear in a memory dump. Finally, we study the inconsistencies introduced by page table smearing in Linux and their impact (Section~\ref{res:smearing}), and we measure the inconsistencies in Linux kernel data structures used by Volatility 2 to extract forensics artifacts (Section~\ref{ref:kernel_structs}).

\section{Idle Kernel Activity}
\label{res:idle_system}
Before analyzing how, and how much, the activity of the kernel introduces inconsistencies during the memory acquisition, we collect some statistics about the interaction of the kernel with the memory in idle state (i.e., when left without user interaction nor external network requests for a sufficiently long interval of time).

For this experiment, we have chosen three OSs: Ubuntu 22.04 and Windows 10 22H2, two of the most used and widely adopted OSs, respectively, in server and desktop environments, and vxWorks 7, a popular real-time OS used in industrial devices and the Internet of Things, but rarely discussed in memory forensics literature. We recorded, using PANDA, the execution trace of an x86\_64 VM equipped with 4GiB of RAM running the three different 64-bit OSs. To assure that the amount of RAM allocated for the VM is sufficient to run the different OSs without incurring effects due to page swapping or race conditions, in which the kernel kills processes to free RAM for new ones, we disabled the swap and, at the same time, ensured that RAM usage in each run never exceeded 75\% of the available memory. It is important to note that systems with larger amounts of installed RAM, even if unused by the operating system, may allocate physical pages causally related more sparsely. This can lead to greater temporal gaps between their dump, potentially increasing the likelihood of inconsistencies. Our choice of dumps size represents therefore a conservative lower bound on the potential inconsistencies that may appear in memory dumps of systems with more RAM.
The selection of RAM size was also driven by a technical trade-off, in addition to the reasons already exposed. We observed a linear correlation between execution time and the memory required to run the custom plugins we developed for PANDA. On a 4 GiB dump, each single measurements can require up to 3 machine days, utilizing 16 cores and up to 256 GiB of RAM limiting the possibility to perform experiments on larger memory dumps.

We booted each machine and left it without any interaction for 30 minutes before performing the experiment to minimize the activity of the OS\footnote{Following preliminary tests we conducted, we observed that after just 15 minutes, the number of memory writes performed by each OS became constant, indicating that the system had entered an idle state.}. We then recorded each kernel for 10 minutes, based on a conservative estimate of the time required by existing tools to acquire 4GiB of memory (as we will describe in more detail in Section~\ref{sec:technique}).

\begin{table*}[t!]
  \scriptsize
    \caption{\label{tab:idle} Statistics about kernels in idle state per
    minute.}
  \scriptsize
  \centering{
  \begin{tabular}{lccccccccc}
      \toprule
      \multicolumn{1}{l}{\textbf{OS}} &
      \multicolumn{1}{c}{\textbf{\thead{Write operations\\ on kernel address \\ space (Millions)}}} &
      \multicolumn{1}{c}{\textbf{\thead{Written Data\\(MiB)}}} &
      \multicolumn{4}{c}{\textbf{\thead{Writes operations\\per size}}} &
      \multicolumn{1}{c}{\textbf{\thead{Unique physical\\pages}}} &
      \multicolumn{1}{c}{\textbf{\thead{Unique virtual\\address}}} \\
      &&&1-byte\hphantom{a}&2-bytes&4-bytes&8-bytes&&\\
      \midrule
      Linux       & 1242 & 8804   & 2.11\% & 0.36\% & 9.98\% & 87.5\% & 129,432 & 60,623,053 \\
      vxWorks     & 29  & 196    & 1.12\% & 0.71\% & 19.14\% & 78.90\% & 145 & 25,209 \\
      Windows 10  & 3885 & 3705  & 6.62\% & 3.55\% & 11.90\% & 77.93\% & 88,784 & 67,248,832 \\
      \bottomrule
  \end{tabular}
  }
  \end{table*}

Table~\ref{tab:idle} summarizes the results. Among the three operating systems, Windows 10 exhibits the highest activity, with approximately 3 times more writing events per minute than Linux and approximately 3 times more data written in terms of quantity. As expected, vxWorks is instead the less active OS, performing only one-thirteenth of the memory operation compared to Windows, probably due to its real-time nature that tends to minimize kernel activity in favor of tasks. Each of the three operating systems demonstrates a notable preference for 8-byte writes (average 81.5\%), possibly due to its optimized use for copying large data quantities. On the other hand, 2-byte writes appear to play a minor role in their operations, accounting for only 1.55\% in average. This suggests that kernel data structures' fields and variables of this size are relatively uncommon. It is also interesting to observe that while Linux writes less often to memory, it touches the highest number of unique physical pages (1.45 times more than Windows). However, due to the higher number of write events per minute, on average, Windows performs more writes per page per minute, approximately 4.5 times more than Linux.

For the Linux kernel, we can classify write events further, as the kernel's virtual address space is divided into regions of fixed sizes~\cite{kerneldocmem}. The majority of writes events (62.03\%) happen in the \texttt{vmalloc} region that contains not-physically continuous pages used to allocate buffers that require to be contiguous in the virtual address space. \texttt{vmalloc} region is used, in particular, to allocate memory for video framebuffers, and this could explain the high number of writes and, at the same time, the low number of unique virtual addresses written (only 0.48\%). 33.28\% of write operations are performed on the direct mapping region that permits the kernel to write directly on every physical page of the system. This region contains the majority of the unique kernel virtual addresses written during the execution (97.35\%). Another 1.67\% of the operations are performed on the virtual memory map area containing the \texttt{struct page} data structures used by the kernel to track the physical pages used in the system. Finally, in the remaining 3.02\% of the events, the kernel writes on its internal global variables, modules, and per-CPU pages.

\newcolumntype{I}{@{\hspace{0.08in}}c}
\newcolumntype{J}{@{\hspace{0.07in}}c}
\newcolumntype{R}{@{\hspace{0.03in}}c}
\newcolumntype{P}{@{\hspace{0.05in}}c}
\newcolumntype{K}{l@{\hspace{1cm}}}
\definecolor{Gray}{gray}{0.8}
\sethlcolor{Gray}

\section{OS Memory Allocation Strategies}
\label{res:other_oss}
The most common approach to obtain a copy of the system memory is to acquire each physical page in ascending order based on its address in the physical address space. This linear dumping strategy is adopted by most software acquisition tools for all major operating systems like Windows and Linux.

To study how different OSs use physical pages and the impact of the linear acquisition technique on the kernel pointers, we have replayed the execution traces of the same VMs used in experiments of Section~\ref{res:idle_system} running a PANDA plugin that emulates the linear acquisition of the memory. Our plugin, by starting from the first physical page available, at a constant rate, saves a hash of the content of each physical page. At the same time, it scans the page by looking for 8-byte values that can be valid kernel virtual addresses. The plugin validates candidate addresses by checking if they are correctly resolved to a physical address by the MMU and if they are part of the canonical range associated with the kernel memory (that for the three OSs in our dataset starts at \texttt{0xffff800000000000}). These addresses represent possible pointers to kernel data structures referenced by some structure contained in the page under dump. Therefore, by saving the hash of the pointed physical page and by comparing it with the one computed when the page is subsequently acquired, we can identify the presence of a pointer inconsistency.

\begin{figure}[t!]
  \includegraphics[width=\columnwidth]{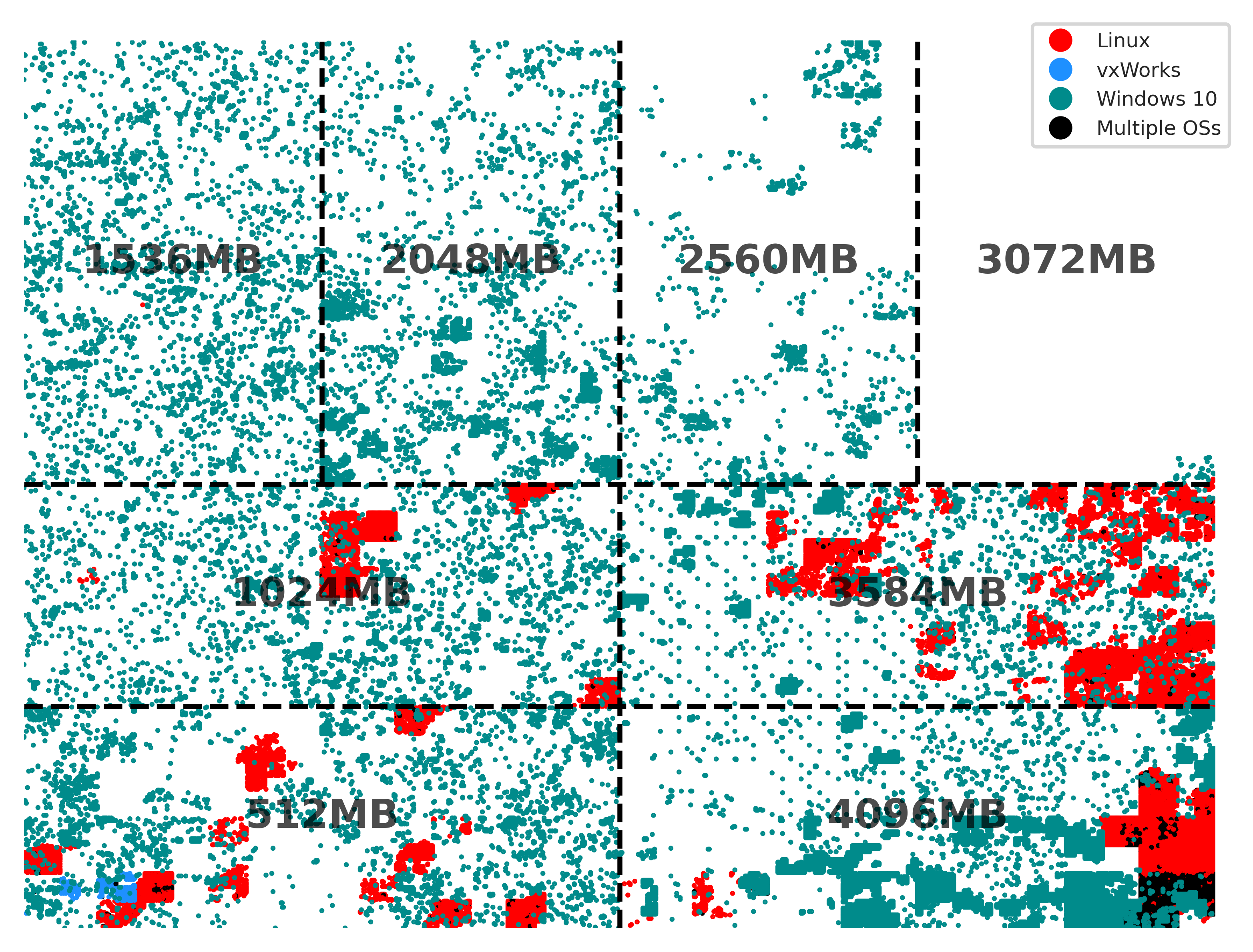}
  \caption{Pages with at least one pointer inconsistency. \label{fig:inconsistencies_hilbert}}
\end{figure}

At the start of the acquisition, our PANDA plugin saves the content of the OS page tables to collect information about how the different OSs manage and map the physical pages onto the kernel virtual address space. Linux and vxWorks, for example, map the entire physical address space by using 2MiB large physical pages in order to directly access each physical page available. This allows them to reduce the number of page table entries needed to address each physical page. When they need to allocate a smaller amount of memory, they generally create a new, separated, mapping by using 4KiB physical pages (sometimes removing the previous 2MiB one). VxWorks tends to create these mappings by using adjacent regions of the physical space, Linux uses more fragmented but still compact regions, while Windows 10 uses large pages to map only a fraction of the entire physical address space and selects random 4KiB pages across the entire physical address space to map smaller regions. On Windows, 47\% of the physical pages that contain a kernel pointer are mapped using only 4KiB mappings, while, on Linux, this value is only 19\% (vxWorks has kernel pointers only in 2MiB pages).

\begin{table}[t!]
  \center
  \caption{\label{tab:pointers_stats} Statistics about kernel pointers
  inconsistencies.}
\centering{
  \scriptsize
\begin{tabular}{lccc}
    \toprule
    \textbf{OS} & \textbf{\thead{Pointer\\inconsistencies\\(absolute)}} &
    \textbf{\thead{Pointer\\inconsistencies\\(relative)}} & \textbf{\thead{Average distance\\among pointers\\(MiB)}}\\
    \midrule
    Linux       & 1,182,656 & 14.6\% & 1120 \\
    vxWorks     & 12,733  & 11.2\% & 14 \\
    Windows 10  & 1,009,818 & 26.7\% & 1740 \\
    \bottomrule
\end{tabular}
}
\end{table}

Table~\ref{tab:pointers_stats} shows the number and ratio of inconsistent pointers in the three OSes. In absolute terms, both Windows and Linux contain over 1 million each, while from a relative point of view, we can see that Windows encounters this problem more frequently (likely due to the much higher frequency at which it updates pointers in memory), resulting in potential inconsistencies in over 25\% of pointers in kernel memory. The last column shows the average distance (in the physical address space) between a pointer and its pointed data. On 4GiB of memory, this distance is 1.1GiB for Linux pointers and 1.8GiB for Windows pointers. For vxWorks, where the memory is all allocated in contiguous chunks, the average distance is instead only 14MB. This is confirmed by Figure~\ref{fig:inconsistencies_hilbert}, which shows a Hilbert plot of the actual pages that contain one or more inconsistent kernel pointers. Linux inconsistencies happen in chunks; Windows ones are spread all over the physical memory, while vxWorks ones are located in a small area on the bottom-left of the figure.

As a reminder, the values provided in this section should be considered as an upper bound on the number of pointer inconsistencies in the acquired memory, as our measurement cannot have false negatives but false positives are possible due to the way we identify candidate pointers (i.e., a large unsigned integer/large negative value that falls within a valid memory range would be considered as a potential pointer).

A more precise measurement would require a detailed knowledge of the OS internals and data structures, which we will present in the following Sections.

\vspace{0.2cm}
\takeaways{Different operating systems adopt very different allocation
strategies, and this choice has a significant impact on the presence of inconsistencies in the
acquired memory. Microsoft Windows is particularly vulnerable to this problem,
which can affect up to a quarter of its kernel pointers. It is also interesting
to note that in Windows the physical distance between the location of a pointer and the location 
of its pointed data is in average 1.8GB (over a 4GB memory), which means that the two
pieces of information are collected \textit{several minutes} apart by the acquisition process.
}

\section{Impact of the Acquisition Technique}
\label{sec:technique}

So far, we have measured the activity and possible inconsistencies of a kernel in an idle state. However, during the acquisition process, the software acquisition tool itself introduces additional noise in the system, as it requires reading and writing a copy of the entire memory to a storage or network device. To measure this impact, and in the rest of the paper, we restrict our focus to the Linux kernel, as it is easier to explore its internals and to modify acquisition tools to collect additional information required by our tests.

\subsubsection*{User vs Kernel-based Acquisition}
The first factor that affects the acquisition noise is whether the acquisition is performed completely in kernel mode or from user space by taking advantage of an already-existent mechanism to access the entire memory of the system. The kernel-based acquisition is much more common, and it is used, for instance, by LiME~\cite{lime}. LiME is a Linux kernel module that dumps the physical memory of a Linux system by copying it to a non-volatile storage or by sending it through the network. After it is loaded, it dumps physical pages by processing one page at a time in ascending order of its physical address. Instead, Microsoft AVML~\cite{avml} is a user-space dumping tool that uses the virtual device \texttt{/proc/kcore} to access the system RAM from the user space. As LiME, AVML dumps the physical pages in ascending order, permitting their saving on a file but managing the entire dump process from the user mode.

We expect user-space solutions to trigger more activity (and, therefore, potentially more inconsistencies) in the kernel. In fact, when running these tools, the kernel has to copy the requested page to user space. The tool then performs some operations (e.g., adding a hash or compressing the data) and then sends the result back to the kernel to write it onto the disk. These additional back-and-forth copies force the kernel to perform more operations, overwrite more pages, and therefore compromise even more the atomicity and efficiency of the acquisition process. To confirm this hypothesis, we recorded the execution trace of a VM Ubuntu 22.04 desktop machine (Kernel 5.19.0-40) equipped with 4GiB of RAM while executing either a LiME or AVML acquisition.

During the execution of LiME, the kernel performed 5576 million write operations, affecting  573.861 different physical pages and a total of 38 GiB of data. The execution of AVML caused instead 9710 million write operations on 796.709 different physical pages, for a total of 68 GiB of data (1,82 more than Lime). As expected, acquiring the memory from user space almost doubles the amount of kernel write operations. As such, this approach should be limited as much as possible, preferring those in kernel space whenever available. Since our purpose is to quantify the inconsistencies that occur in a dump under the best possible experimental conditions, we will use LiME as a memory acquisition tool for the rest of our experiments.

\subsubsection*{Storage}
We now look at the impact of the target storage that is used to record the acquired memory image. In particular, this covers the interplay of different aspects, including where the image is saved (network vs. internal disk vs. USB disk), how (by using kernel functions or direct access to the device), and which filesystem is used to store it.

LiME has three main operating modes that can alter the content of the memory during the dump process:

\vspace{0.2cm}
\noindent \textbf{Dump on disk} -- In this mode, LiME uses kernel primitives to write the memory content on a file. The specific type of target hard drive, an internal SATA drive or an external USB drive, can impact the number of physical pages marked as "dirty" by the kernel during the dumping process. This is because the kernel copies the content of a page being dumped into various caches, including bus and device drivers, ring buffers, page cache, and others. Furthermore, different target filesystems use, internally, different algorithms, caches, and data organization, resulting in different memory footprints generated during the writing operations.

\vspace{0.2cm}
\noindent \textbf{Dump on disk using Direct I/O} -- In this mode, LiME opens the target file using the \texttt{O\_DIRECT} flag, permitting the bypass of the kernel's page cache, reducing the number of physical pages dirty. However, direct I/O is required to be supported by the underlying filesystem.

\vspace{0.2cm}
\noindent \textbf{Dump through the network} -- In this mode, LiME sends the content of the RAM through a TCP connection established by an external system.

To reduce the impact of kernel writes on memory due to user space disk and network activity, and measure specifically the effect of the different LiME dump modes, we use a minimal VM running Linux equipped with 1 GiB of RAM that runs only two userspace processes: \texttt{getty} and \texttt{sh}. After waiting for the kernel to enter the idle state, we dump the memory on an emulated USB external hard drive and on an emulated internal SATA drive. We tested 8 commonly used file systems in buffered and direct I/O mode, as well as, dump through the network mode. LiME has been loaded with the options \texttt{timeout = 0} to disable the default timeout, beyond which LiME ignores the page, and \texttt{format = raw}, which instructs LiME to copy the memory as is without adding LiME file format headers in order to enable \texttt{O\_DIRECT}.

Table~\ref{tab:modes} presents the measurement results conducted on file systems using an emulated external USB disk, an internal SATA disk, and a network dump. In each column, the highlighted value is the best result among the test cases of the same type. The most significant indicator to determine how much the dump method dirties the memory content is the number of different physical pages written during the dump. As we can see in Column 5, for dump performed using buffered I/O, there are no relevant differences, for the same file system, between the number of physical pages written on a USB disk or SATA one (a maximum difference of 0.23\%) and, as well as, there are no relevant differences in the number of pages written between file systems tested (a maximum difference of ~1\%). However, different file systems perform differently in terms of the time needed to complete the dump and the number of writing events. In particular, on a USB disk, XFS requires ~45\% less time and 56\% less writing operations than the widely adopted FAT32 file system. These differences can have an impact on the number of inconsistencies inside a memory dump because the more time and operations are required to complete the dump process, the higher the probability for the kernel to write on pointers located on pages already dumped.

Only 3 file systems in our tests (Btrfs, exFAT, and FAT32) support direct I/O in kernel mode. In this case, as we can see from Column 3, the number of writes on MMIO regions is increased by a factor of 100, a sign that the kernel has changed the method to access the physical device to directly write on them. In all three file systems, we also observe a drastic reduction, up to 99.8\%, in the number of different physical pages written during the dump process. This very positive aspect is, however, counterbalanced by a dramatic increase in the time required to acquire the memory, which increased by up to 100 times. To understand the reasons for this huge increase in dump time, we have analyzed calls to the allocation/free functions used internally by the kernel to manage disk write operations. During our investigation, we found that for every page written on disk, the Block I/O subsystem of the kernel allocates and releases a \texttt{struct bio} along with its associated \texttt{mempool}. The memory for these operations is obtained from the \texttt{bio-160} memory pool, which, in turn, relies on the kernel SLAB memory pool system. All of these memory operations significantly contribute to an increase in the dump time and the quantity of data written in terms of metadata. Furthermore, unlike buffered I/O, which optimizes disk writes by using the page cache and adapts to the underlying hardware, direct I/O mode can reveal differences due to the distinct interfaces involved. We have observed a relevant disparity in the time required to perform a dump on a USB drive compared to a SATA drive. This discrepancy is not limited to PANDA artifacts, as we have also observed it on real hardware. It can be attributed to the different implementations of the USB and SATA kernel subsystems, as well as the varying complexity and number of layers in their respective protocol stacks.

Finally, the network dump is the most balanced method in terms of physical pages modified (only 22 more than the fastest direct I/O method), amount of time needed (1.5x the fastest buffered I/O file system), total number of events, and total size of data written.

\vspace{0.2cm}
\takeaways{Our experiments demonstrate that for Linux bare-metal systems
the most efficient software-based approach to acquire a complete memory
dump—while minimizing the number of physical pages written during the
dumping process—is to use a kernel module that reads system memory and
transmits its contents over the network. 
When this is not possible,
we recommend adopting a traditional method by saving the memory to an
external disk formatted with an \texttt{EXT} or \texttt{XFS} filesystems.
External disks formatted with the \texttt{FAT32} system, unfortunately a very popular
choice among forensic analysts, provide instead the worst performances.}
\vspace{0.2cm}

\begin{table*}[t!]
  \caption{\label{tab:modes} Statistics about the dump process in different
  conditions.}

\scriptsize
\centering{
\begin{tabular}{lcccccccccccc}
  \midrule
  \multicolumn{1}{l}{\textbf{Mode}} &  \multicolumn{2}{c}{\thead{\textbf{Writes operations on}\\ \textbf{kernel address space} \\ \textbf{(Millions)} }} & \multicolumn{2}{c}{{\thead{\textbf{
    Writes on} \\\textbf{MMIO regions}}}} & \multicolumn{2}{c}{\thead{\textbf{Total size} \\ \textbf{(GiB)}}} &
    \multicolumn{2}{c}{\thead{\textbf{Unique physical}\\ \textbf{pages}}} &
    \multicolumn{2}{c}{\thead{\textbf{Time required}\\\textbf{(ratio)}}} \\
    & USB & SATA & USB & SATA & USB & SATA & USB & SATA & USB & SATA \\
    \midrule
    Btrfs &    874 & 811 &  59824 & 37778 &   6.01 & 5.58 &  \hl{249340} & 249204 &    1.81x & 1.53x \\
    exFAT &   1005 & 938 &  96112 & 61772 &   6.89 & 6.43 &  251074 & 250728 &    1.79x & 1.33x \\
    Ext4  &    818 & 757 & 60692 & 35696 &     5.61 & 5.20 &  249421 & 248873 &    1.76x & 1.16x\\
    Ext4 no journal & 776 & 719  &  61744 & 36443 &  5.33 & 4.95 &  249439 & 248864 &    1.78x & 1.10x\\
    F2FS  &    951 & 910 &  61329 & 36743 &   6.51 & 6.23 &  249406 & 249379 &    2.04x & 1.33x\\
    NTFS  &    796 & 739 &  61329 & 38711 &    5.48 & 5.09 &  249411 & 249129 &    1.75x & 1.31x\\
    FAT32 &   1404 & 1317 &  84456 & 89542 &   9.65 & 9.06 &  250328 & 250908 &    2.39x & 1.82x\\
    XFS   &  \hl{632} & \hl{569} &  \hl{57605} & \hl{34255} & \hl{4.41} & \hl{3.97} &  249405 & \hl{249041} &    \hl{1.62x} & \hl{1x}\\
    \midrule
    Btrfs D. I/O &  49137 & 37708 &  9147061 & 6707022 & 344.04 & 265.19 &   75037 & 78885 &  255.76x & 73.00x\\
    exFAT D. I/O &  \hl{10698} & \hl{4713} &  \hl{5204034} & \hl{3950081} &  \hl{73.20} & \hl{32.11} &  \hl{466} & \hl{497} & \hl{109.34x} & \hl{15.85x} \\
    FAT32 D. I/O &  16657 & 6000 &  9138277 & 5773820 &  114.15 & 40.88 & 1125 & 1127 &  236.71x & 19.58x \\
    \midrule
    Network &  \multicolumn{2}{c}{1336} & \multicolumn{2}{c}{1000453}& \multicolumn{2}{c}{8.64} & \multicolumn{2}{c}{488} &  \multicolumn{2}{c}{2.73x}\\
    \bottomrule
\end{tabular}
}

\end{table*}

\section{Types of Inconsistencies}
\label{sec:approach_inconsistencies}

\begin{figure}[h!]
  \centering {
    \includegraphics[width=0.8\columnwidth]{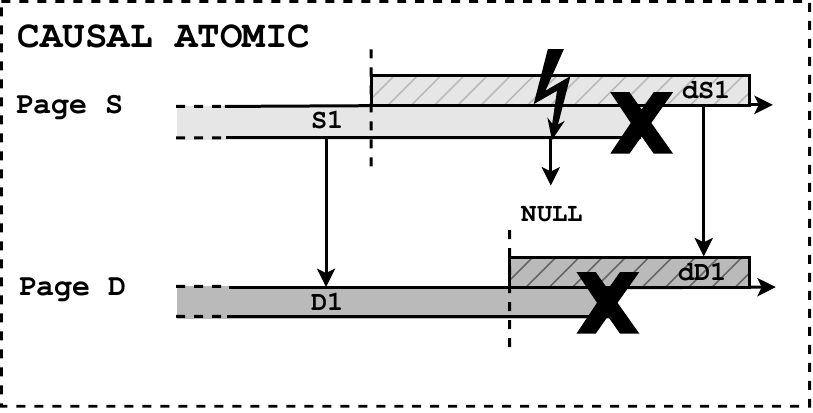}
  \caption[Example of causal atomic dump]{\label{fig:good_dump} Example of
  causal atomic dump.}
  }
\end{figure}

\begin{figure}[h!]
  \centering{
    \includegraphics[width=0.8\columnwidth]{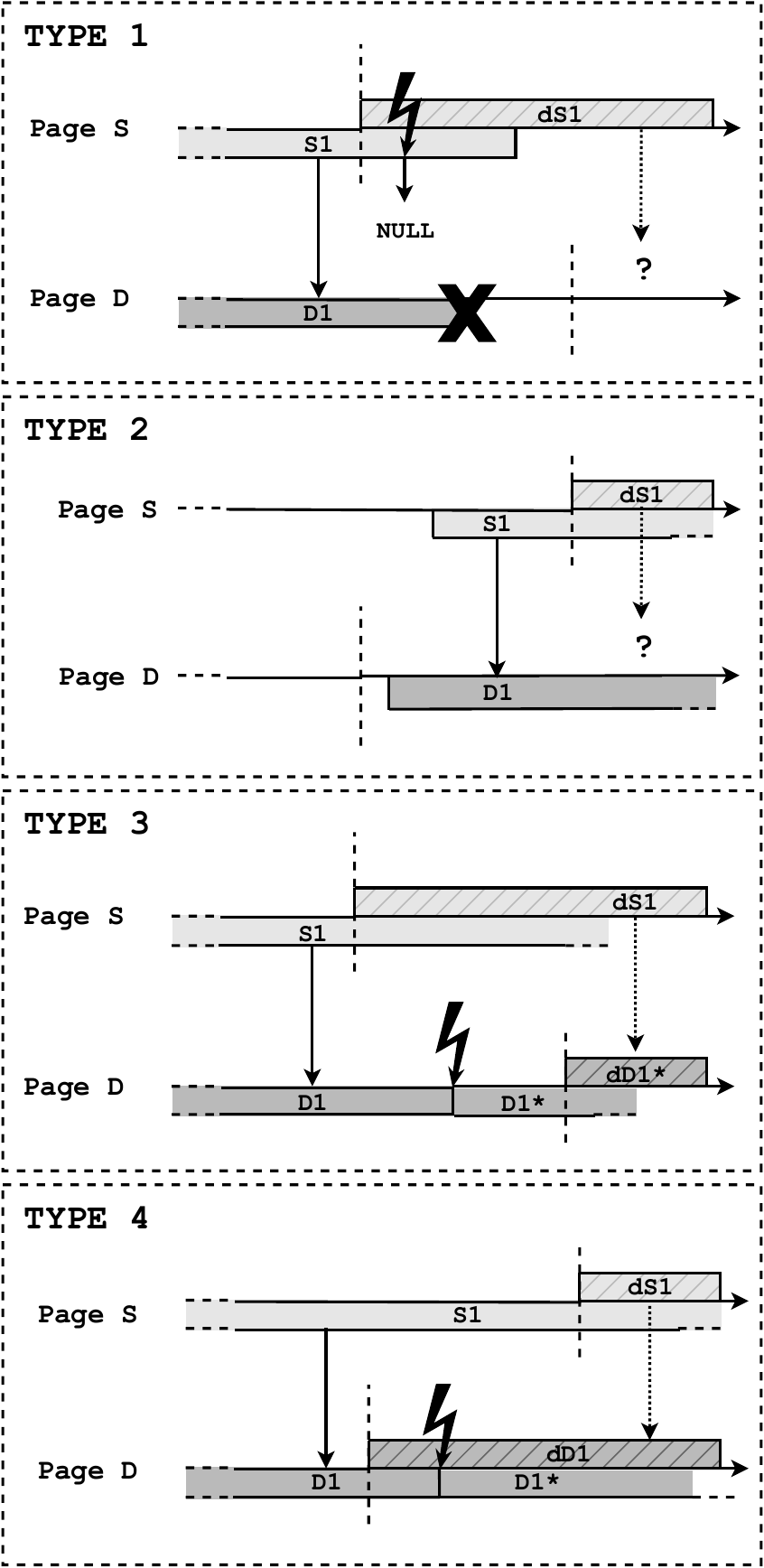}
    \caption[Types of inconsistencies]{\label{fig:inconsistencies} Types of
    inconsistencies.}
  }
\end{figure}

So far, we have discussed possible inconsistencies in relation to kernel pointers that connect two physical pages. However, in the case of structures of known sizes, such as page tables and kernel data structures, we can perform a more in-depth analysis and classification of different types of inconsistencies. In our measurement, we consider five different types of inconsistencies that affect memory structures. Four of them are related to the order and time difference between the acquisition of the pages that contain two different causally-related structures. The last is instead caused by a timing difference in the dump of two or more pages that contain fragments of a single large structure. To visualize the various inconsistencies, we will use a series of diagrams similar to the one shown in Figure~\ref{fig:good_dump}. The diagram shows two structures \texttt{S1} and \texttt{D1} entirely contained in physical pages \texttt{S} and \texttt{D}. The horizontal axis of the diagram represents time. Structure \texttt{D1} is referenced by structure \texttt{S1} through a pointer, depicted by a vertical arrow on the left side of the figure.

At a specific moment, indicated by a vertical dashed line, the dump tool saves the contents of page \texttt{S}, and consequently, the contents of structure \texttt{S1}, to the dump file. From this point on, in Figure~\ref{fig:good_dump}, there are two bands associated with the structure \texttt{S1}: the lower band, which continues to represent the structure \texttt{S1} in memory, and the upper knurled band \texttt{dS1}, which represents the unmodifiable copy of structure \texttt{S1} saved within the dump file. After a while, also the page \texttt{D} is dumped, resulting in a split of the associated band. After the dump of page \texttt{D}, the kernel performs a memory write operation, represented as a thunderbolt in Figure~\ref{fig:good_dump}, invalidating the reference to structure \texttt{D1}. Finally, the kernel deallocates both structures, an operation represented as a cross in the diagram.

As shown in the right part of the figure \texttt{dS1}, the dumped copy of \texttt{S1} structure correctly contains a reference to \texttt{dD1}, the dumped copy of \texttt{D1} structure: the dump of the two structures and their relation is causal atomic since the causal relationship between the two structures is preserved by the dump.

We can classify inconsistencies into two categories: causal inconsistencies and value inconsistencies. In causal inconsistencies, the causal relation between structure \texttt{S1} and structure \texttt{D1} is compromised, and the structure \texttt{dS1} in the dump file does not point to the correct structure \texttt{dD1}, but to generic data that can be located at the same address. In value inconsistencies, instead, the causal relation among structures is preserved, but the content of the pointed structure \texttt{D1} is changed during the dump process. Below, we detail the five types of inconsistencies, four of which are illustrated in Figure~\ref{fig:inconsistencies}:

\vspace{0.2cm}
\noindent \textbf{Type 1 (causal inconsistency)} -- The kernel de-allocates the destination structure \texttt{D1} after dumping \texttt{S1} but before dumping \texttt{D1}, potentially replacing the memory with different data. As a result, the dumped \texttt{dS1} structure references unrelated data, and the dump does not contain the original content of structure \texttt{D1}.

\vspace{0.2cm}
\noindent \textbf{Type 2 (causal inconsistency)} -- The kernel allocates a couple of structures, with the referenced one on an already dumped page. It is the mirror case of Type 1 when the destination structure \texttt{D1} is dumped before \texttt{S1}.

\vspace{0.2cm}
\noindent \textbf{Type 3 (value inconsistencies)} -- The kernel modifies the content of the pointed structure before it is acquired but after the dump of \texttt{S1} structure: \texttt{dS1} in the dump will reference the modified version of \texttt{D1} instead of the original one. This inconsistency can also arise when, after discarding structure \texttt{S1}, the kernel frees up structure \texttt{D1} and substitutes it with another structure of the same kind that has no causal connection to \texttt{S1}. In this scenario, within the dumped data, a structure of the same type as \texttt{D1} exists, but it is impossible for the analyst to determine whether it represents a modified version of \texttt{D1} or a subsequent allocation. An example of such a case can occur in Linux if the \texttt{D1} structure is a part of a kernel SLAB.

\vspace{0.2cm}
\noindent \textbf{Type 4 (value inconsistencies)} -- It is the mirror case of Type 3 when the destination structure \texttt{D1} is dumped before \texttt{S1}.

\vspace{0.2cm}
\noindent \textbf{Type 5 (value inconsistencies)} -- This inconsistency, not represented in Figure~\ref{fig:inconsistencies}, occurs when a structure occupies more than one physical page or it is on the fence between two. In this scenario, after the dump of a page that contains a portion of the structure, the kernel modifies one of the other pages composing it: the version of the structure saved on the dump will then consist of fragments obtained at different points in time. As a result, the values within the dumped structure will not be globally consistent.

\begin{table}
  \caption{\label{tab:page_amomalies} Total number of inconsistencies per type in page tables.}

\centering{
\scriptsize

\begin{tabularx}{1\columnwidth}{XIIIIIIIIII}
    \toprule
     &  $\mathbf{D_0}$ & $\mathbf{D_1}$ & $\mathbf{D_2}$ &
    $\mathbf{D_3}$ & $\mathbf{D_4}$ & $\mathbf{D_5}$ & $\mathbf{D_6}$ & $\mathbf{D_7}$ &
    $\mathbf{D_8}$ & $\mathbf{D_9}$\\
    \midrule
    Type 1  & - & - & - & - & 35 & - & 302 & - & 1 & -\\
    Type 2  & 1 & - & - & 3 & 22 & 6 & 45 & - & 63 & -\\
    \midrule
    Type 3  & - & 3 & 6 & 33 & 57 & 26 & 198 & - & 60 & 22\\
    Type 4  & 2 & 3 & 5 & 21 & 24 & 40 & 24 & 10 & 23 & 21\\
    \bottomrule

\end{tabularx}
}
\end{table}

\begin{table}
  \caption{\label{tab:proc_anomalies} Processes with inconsistencies in private page tables.}
\centering{
  \scriptsize
  \begin{tabularx}{1\columnwidth}{XIIIIIIIIII}
    \toprule
    &  $\mathbf{D_0}$ & ${\mathbf{D_1}}$ & $\mathbf{D_2}$ &
    $\mathbf{D_3}$ & $\mathbf{D_4}$ & $\mathbf{D_5}$ & $\mathbf{D_6}$ & $\mathbf{D_7}$ &
    $\mathbf{D_8}$ & $\mathbf{D_9}$\\
    \midrule
    Unique processes & 2 & 2 & 4 & 11 & 14 & 12 & 26 & 3 & 24 & 19\\
    \midrule
    Type 1 & - & - & - & - & 4 & - & 19 & - & - & -\\
    Type 2 & 1 & - & - & - & 3 & 1 & 2 & - & 3 & -\\
    \midrule
    Type 3 & - & 2 & 2 & 9 & 12 & 9 & 18 & - & 17 & 11\\
    Type 4 & 2 & 2 & 3 & 6 & 6 & 10 & 4 & 3 & 9 & 11\\
    \bottomrule
\end{tabularx}
}

\end{table}

\begin{table}[t!]
  \caption{\label{tab:kern_anomalies} Dumps with at least a kernel page table with inconsistencies.}
\centering{
  \scriptsize
  \begin{tabularx}{1\columnwidth}{XIIIIIIIIII}
    \toprule
     &  $\mathbf{D_0}$ & ${\mathbf{D_1}}$ & $\mathbf{D_2}$ &
    $\mathbf{D_3}$ & $\mathbf{D_4}$ & $\mathbf{D_5}$ & $\mathbf{D_6}$ & $\mathbf{D_7}$ &
    $\mathbf{D_8}$ & $\mathbf{D_9}$\\
    \midrule
    Type 1  &   &   &   &   &   &   &   &   &   &  \\
    Type 2  &   &   &   & \checkmark &   &   & \checkmark &   &   &  \\
    \midrule
    Type 3  &   & \checkmark &   & \checkmark & \checkmark & \checkmark & \checkmark &   &   &  \\
    Type 4  &   &   &   & \checkmark & \checkmark &   & \checkmark &   & \checkmark &  \\
    \bottomrule
\end{tabularx}
}
\end{table}

\section{Page Table Smearing}
\label{res:smearing}

The first type of memory structure that we study in our experiments is page tables, as page table smearing (i.e., the presence of inconsistent or malformed page tables in memory dumps) is often reported as one of the main problems of non-atomic acquisitions.

To study this phenomenon, we developed a PANDA plugin that tracks the creation/destruction of radix trees in the system\footnote{Intercepting calls to functions \texttt{pgd\_ctor} and \texttt{pgd\_free}.}, allocations and deallocations of page tables of lower-levels\footnote{Intercepting calls to functions \texttt{\_\_\_pte\_free\_tlb}, \texttt{\_\_\_pmd\_free\_tlb}, \texttt{\_\_\_pud\_free\_tlb}, \texttt{pud\_free\_pmd\_page}, \texttt{\_\_free\_pages\_ok} and \texttt{free\_unref\_page}.}, and all the kernel writes. We performed our analysis on memory dumps of a virtual machine runs Ubuntu 22.04 desktop (Kernel 5.19.0-40) equipped with 4GiB of RAM and repeated each experiment 10 times to reduce the effect of randomness.

For each run, we booted the machine and started a set of commonly used applications. After 30 minutes of inactivity, the kernel and user applications consumed about 25\% of the available memory. Then, we started a memory acquisition using LiME and dumped the memory on an emulated external USB drive formatted with the Ext4 file system. We decided on this configuration because an external disk allows the analyst to preserve the content of the hard disk of the target machine (often a requirement in an investigation) and because Ext4 (see Table~\ref{tab:modes}) introduces a minimal overhead compared with the other filesystem (e.g., FAT32) whose support is compiled by default in the kernels of all major distributions.

The running applications are chosen to simulate an office environment and include: \texttt{firefox} with 6 tabs opened, \texttt{libreoffice}, the \texttt{thunderbird} E-Mail client, and the \texttt{evince} PDF reader. Before the start of the acquisition, we counted an average of 210 processes running on the system. The LiME acquisition took $9.18\pm5.46$ minutes to complete (on 5 dumps, it requires less than 5 minutes while in one case, Dump 6, up to 17 minutes). This variability in the acquisition time could derive from the activity of the user space processes which can slow down the dump and require the kernel to perform privileged operations for them. LiME does not use any mechanism to prioritize its activity, such as starting a high-priority kernel task to dump the memory from.

Table~\ref{tab:page_amomalies} summarizes the total number of inconsistencies, categorized by type, for each of the 10 memory dumps. All dumps exhibit inconsistencies, ranging from a minimum of 3 (Dump 0) to a maximum of 569 (Dump 6). Table~\ref{tab:proc_anomalies} details the number of individual processes in each dump affected by at least one inconsistency in the private part of their address spaces (i.e., address spaces not containing kernel data or code) and the number of processes impacted by each specific type of inconsistency. A comparison between the first row of the table and the subsequent ones for each memory dump reveals that multiple types of inconsistencies can simultaneously affect the same process. 

Additionally, Table~\ref{tab:kern_anomalies} highlights dumps that contain at least one inconsistency in the kernel page tables. It is important to note that, on the Intel architecture, if an inconsistency exists in a kernel page table, it will appear in the radix trees of all processes. This occurs because the kernel page tables are included in the radix tree of every process in the system. In our dataset, we observed this phenomenon in 60\% of the memory dumps.  It may also occur that inconsistencies exist only in kernel page tables part of a process radix tree but not in the process private one, as seen for Type 2 inconsistencies in Dump 3. 

Furthermore, the level at which an inconsistency occurs within the radix tree -- whether in the page table corresponding to the user or kernel address space -- directly determines the extent of its impact on the virtual address space. Specifically, inconsistencies closer to the root of the tree affect larger portions of the address space. Across the 10 memory dumps analyzed, we identified 91 inconsistencies in entries of the top-level page tables (level 0, closest to the root), 25 at level 1, and 940 at level 2.

For inconsistencies of Types 2 and 3, we can distinguish two cases: when the affected page table is only modified by the kernel during the acquisition and when it is instead deallocated and substituted by another one.
In the first case, our PANDA plugin also permits quantifying exactly how much of the private virtual address space of each process is affected by the anomaly: on average, $770$ KiB (with a standard deviation of $2.85$ MiB) and up to a maximum of $64.63$ MiB present anomalies such as missing pages, wrong permissions or errors in translation. In this case, the analyst might miss pieces of evidence because she cannot fully explore the user-space processes. In the second case, things are even worse, as the page tables of one process can erroneously point to the page tables of another one. As a result, the address space of the first process will contain and refer to data belonging to the second process without the analyst being able to tell the difference. This virtual address space anomaly can potentially deceive analysts in many ways. For instance, it can create the illusion that pieces of code from one process exist within the address space of another, implying potential manipulation or injection of code. Moreover, it can lead to a scenario where sections of one process's heap are replaced by fragments from another, rendering the analysis of the process's data structure unattainable. As rare and unlikely as this may seem, in our dataset we have found two dumps, $D_4$ and $D_6$, in which this anomaly occurs. In particular, in $D_6$, two of the office applications that we have started, (\texttt{libreoffice} and \texttt{thunderbird}) have erroneously attributed memory pages that belong to the graphical environment manager (\texttt{gdm}).

To make things worse, while in theory (even though for the best of our knowledge, no tools attempt to do so), one could try to detect, but not correct, causal inconsistencies (Type 1 and Type 2) the presence of a value inconsistency (Type 3 and 4) cannot be recognized nor corrected. In fact, the presence of a false page table introduced by a causal inconsistency could be detected by employing a validation model that checks the page tables against the inviolable constraints defined by the MMU, as recently discussed in~\cite{mmushell}. However, in cases of value inconsistency, the erroneous page tables introduced may still appear valid as the automatic recognition of such anomalies cannot identify them, and even a human analyst may encounter difficulties in identifying them as inconsistencies.

\vspace{0.2cm}
\takeaways{Our experiments reveal that page smearing is a very
common problem, affecting user-space processes in 100\% of the dumps in our
dataset and affecting kernel memory in 60\% of them. The extent of anomalies in a process's
address space depends directly on the level of the page table radix-tree
where the inconsistency occurs. In our dataset, we observed a case where up
to 64 MiB of a process's private address space was affected by anomalies.
Additionally, we show that a single process can simultaneously exhibit
multiple types of inconsistencies in its private address space, many of
which are difficult—if not impossible—to detect during the analysis. Furthermore, we
found cases in which these inconsistencies led to portions of memory 
from different processes being mixed together or erroneously assigned to the wrong process.}

\section{Kernel Structures}
\label{ref:kernel_structs}

\begin{table}
\caption{\label{tab:struct_amomalies} Total number of inconsistencies by type in kernel
data structures.}
\centering{
  \scriptsize
  \begin{tabularx}{1\columnwidth}{XIIIIIIIIII}
    \toprule
     &  $\mathbf{D_0}$ & $\mathbf{D_1}$ & $\mathbf{D_2}$ &
    $\mathbf{D_3}$ & $\mathbf{D_4}$ & $\mathbf{D_5}$ & $\mathbf{D_6}$ & $\mathbf{D_7}$ &
    $\mathbf{D_8}$ & $\mathbf{D_9}$\\
    \midrule
    Type 1  & 22 & 37 & 6 & 16 & 5936 & 16 & 4733 & 7 & 131 & 7\\
    Type 2  & 10 & - & 2 & 135 & 1455 & 11 & 1546 & - & 2227 & 7\\
    \midrule
    Type 3  & 1773 & 1740 & 670 & 3496 & 9772 & 3551 & 4040 & 570 & 1400 & 597\\
    Type 4  & 1961 & 2010 & 1272 & 5315 & 4245 & 4733 & 3183 & 1225 & 1981 & 1321\\
    \bottomrule

\end{tabularx}
}

\end{table}

\begin{table*}
  \caption{\label{tab:struct_type_stats} Average number of inconsistencies per
  kernel structure over the 10 dumps.}

\centering{
\scriptsize
\begin{threeparttable}
\begin{tabular}{lcccccccc}
    \toprule
    \textbf{Struct Type} & \textbf{\thead{Struct Size\\(bytes)}} & \textbf{\thead{Number\\of\\Instances}} &
    \multicolumn{4}{c}{\textbf{Inconsistencies}} & \textbf{\thead{Percentage \\with\\ inconsistencies}}\\
    &  & & Type 1 & Type 2 & Type 3 & Type 4 & \\
    \midrule
    \texttt{cred} & 176 & 2248 & - & - & - & - & - \\
    \texttt{dentry} & 192 & 322,248 & 687 & 2630 & 13,445 & 16,686 & 7.36\% \\
    \texttt{dentry\_operations} & 128 & 118 & - & - & - & - & - \\
    \texttt{fdtable} & 1188 & 1108 & - & 6 & 34 & 127 & 15.07\% \\
    \texttt{fdtable array}\tnote{1} & - & 1107 & 21 & 67 & 59 & 148 & 19.42\% \\
    \texttt{file} & 232 & 80,302 & 243 & 151 & 218 & 207 & 1.02\% \\
    \texttt{file\_system\_type} & 72 & 402 & - & - & - & - & - \\
    \texttt{files\_struct} & 704 & 1108 & - & - & 9 & - & 0.81\% \\
    \texttt{fs\_struct} & 56 & 1134 & - & 2 & 61 & - & 5.56\% \\
    \texttt{inode} & 632 & 253,983 & - & 40 & - & - & 0.02\% \\
    \texttt{mm\_struct} & 1048 & 1104 & 10 & 17 & 14 & 10 & 4.08\% \\
    \texttt{sock} & 760 & 7302 & - & - & - & - & - \\
    \texttt{socket\_alloc} & 768 & 7210 & 35 & 52 & 19 & 41 & 2.04\% \\
    \texttt{super\_block} & 1472 & 236 & - & 1 & - & 10 & 4.66\% \\
    \texttt{task\_struct} & 9792 & 2132 & 62 & 29 & 312 & 412 & 22.56\% \\
    \texttt{vfsmount} & 32 & 488 & - & 2 & 8 & 18 & 5.74\% \\
    \texttt{vm\_area\_struct} & 208 & 348,551 & 9853 & 2396 & 13,430 & 9587 & 9.19\% \\
    \bottomrule
\end{tabular}
\begin{tablenotes}
\item[1] This is a variable size array of pointers to \texttt{struct files} pointed by \texttt{fdtable.fd} field.
\end{tablenotes}
\end{threeparttable}
}
\end{table*}

\begin{table}[h!]
  \centering
  \scriptsize
  \caption{\label{tab:fields} Fields with at least one inconsistency in top 5 structures in terms of percentage of affected instances.}
\begin{tabularx}{\columnwidth}{@{}p{2cm}p{1.8cm}|p{1.8cm}p{1.7cm}}
  \toprule
  \textbf{Structure} & \textbf{Field Name} & \textbf{Structure} & \textbf{Field Name}\\
  \midrule
  \texttt{task\_struct} & \texttt{children.next} & \texttt{task\_struct} & \texttt{children.prev}\\
  \texttt{task\_struct} & \texttt{comm} &\texttt{task\_struct} & \texttt{cred}\\
  \texttt{task\_struct} & \texttt{files} & \texttt{task\_struct} & \texttt{parent} \\
  \texttt{task\_struct} & \texttt{sibling.next} & \texttt{task\_struct} & \texttt{sibling.prev} \\
  \texttt{task\_struct} & \texttt{task.next} & \texttt{task\_struct} & \texttt{task.prev}\\
  \texttt{task\_struct} & \texttt{thread\_group}\\
  \midrule
  \texttt{fdtable array} & Array entries & \texttt{fdtable} & \texttt{fd}\\
  \midrule
  \texttt{vm\_area\_struct} & \texttt{vm\_end} & \texttt{vm\_area\_struct} & \texttt{vm\_file}\\
  \texttt{vm\_area\_struct} & \texttt{vm\_flags} & \texttt{vm\_area\_struct} & \texttt{vm\_mm}\\
  \texttt{vm\_area\_struct} & \texttt{vm\_next} & \texttt{vm\_area\_struct} & \texttt{vm\_pgoff}\\
  \texttt{vm\_area\_struct} & \texttt{vm\_rb.rb\_left} & \texttt{vm\_area\_struct} & \texttt{vm\_rb.rb\_right}\\
  \texttt{vm\_area\_struct} & \texttt{vm\_start}\\
  \midrule
  \texttt{dentry} & \texttt{d\_child} & \texttt{dentry} & \texttt{d\_inode}\\
  \texttt{dentry} & \texttt{d\_name.name} & \texttt{dentry} & \texttt{d\_op}\\
  \texttt{dentry} & \texttt{d\_subdirs}\\
  \bottomrule
  \end{tabularx}
\end{table}

In our previous set of experiments, we observed that memory acquired in a non-atomic way contains a large number of inconsistencies among pointers and page table entries. We now focus on those inconsistencies that affect fields of kernel data structures that are commonly used by existing tools for evidence recovery from memory dumps.

In particular, we consider as forensics-relevant data structures all those used by at least one of the base set of Volatility 2 plugins. However, some of them that contain very important forensic artifacts are rarely modified during a normal execution. For instance, the \texttt{modules} linked list (composed by \texttt{struct module} nodes) maintains information about kernel modules loaded in the system and, therefore, only experiences modifications when a module is loaded or unloaded from the system, an operation that rarely happens during the memory acquisition process. Therefore, non-atomic inconsistencies among these structures are exceptionally rare in practice. For this reason, we will focus our analysis on those forensic relevant structures that are routinely modified by a running kernel.

In particular, among those traversed and explored by Volatility 2's plugins, we selected 17 important kernel data structures that the OS frequently modifies. A complete list is shown in Column 1 of Table~\ref{tab:struct_type_stats}. These structures maintain highly volatile forensics-relevant information about the running processes, their resources, opened files and sockets, and part of the file system information cached in memory.

To quantify the presence of inconsistencies in these structures, we created another PANDA plugin that tracks the evolution of the content and links among kernel data structures during an acquisition process. Before the start of the dump, the plugin explores the memory content of the VM, identifying kernel data structures and dereferencing pointers contained in them. Then, during the replay of the execution trace, it intercepts allocations of new data structures, deallocations, and modifications of already tracked ones by maintaining information about the version of each structure as already explained in Section~\ref{sec:measure_technique}.

For each of the 17 different structure types, we track separately two types of fields: the pointer fields that maintain relations among data structures and that are traversed by Volatility plugins to reach other structures, and the data fields that contain the information retrieved by the plugins. In total, we tracked 38 pointers and 40 data fields in our experiments. It is important to note that this number corresponds to the subset of fields used by Volatility 2 and not all the fields in the data structures since many have no forensics relevance.

Table~\ref{tab:struct_amomalies} reports the number of data structures affected by at least one inconsistency in each of the ten images. As in the case of page smearing, each dump contained at least one value inconsistency (Type 3 and 4) and a causal inconsistency of Type 1, ranging from 1950 to over 20K inconsistencies.

Again, as in the case of page smearing, dumps $D_4$ and $D_6$ had the highest number of structures with inconsistencies. At a closer look, these two dumps contain the largest total number of tracked structures allocated and deallocated by the kernel during the dump process, with 275,830 and 327,907, respectively (the average is $185,151$). This suggests that the kernel was more `active' during the acquisition process. This additional kernel activity can be attributed to various factors that occur during the system's normal operation, even in an idle state. For example, memory allocation and opening/closing file and network handles by system daemons, the launch of scheduled processes like automatic system update managers, or internal kernel optimizations, such as flushing page caches.

In our kernel configuration, there are 9,593 different types of kernel data structures\footnote{These can be counted using the \texttt{pahole} utility.}, of which only 103 (1.07\%) have a size greater than a physical page. Among those tracked in this experiment, only the \texttt{task\_struct} falls into this category, with a size of 9,792 bytes. Furthermore, our analysis of the dump dataset revealed that less than 0.1\% of tracked data structures smaller than a page are allocated across two different physical pages. These findings make Type 5 inconsistencies an extremely rare event among forensically relevant structures, and in fact, none were encountered in our experiments.

To better understand how the inconsistencies affect the different types of kernel data structures in Table~\ref{tab:struct_type_stats}, we report the size of each structure, the mean value of the number of structures present in a dump, the mean value of the number of data structures affected by inconsistencies and the ratio between the mean number of unique structures affected and the mean number of structures of that type. The four more common structures (\texttt{vm\_area\_struct}, \texttt{dentry}, \texttt{inode}, \texttt{file}) account for over 98\% of the structures we track in memory and contain 98\% of the inconsistencies. Unfortunately, these are also the most used by Volatility 2 plugins, as we will discuss in more detail later in this section. For example, inconsistencies in data structure related to cached information about physical and virtual filesystems such as \texttt{dentry} and \texttt{inode} can compromise the ability to extract files from volatile filesystems like \texttt{/tmp}. Files in \texttt{tmpfs} are not saved on disk but reside solely in RAM, making their recovery challenging when these data structures are compromised. It is also important to note that while there are relatively few instances of \texttt{task\_struct}, they contain the highest number of inconsistencies among the traced data structures (22\%). In particular, they are affected by a large number of value inconsistencies: this means that dereferencing a traced pointer in a \texttt{task\_struct}, the pointed structure has, roughly, a 1 in 4 chance to contain information not in sync with those of the \texttt{task\_struct} itself. This can mislead analysts in diverse ways. For instance, if the inconsistency impacts the \texttt{cred} pointer, a process may falsely exhibit lower privileges than it truly possessed. Similarly, if the inconsistency affects the \texttt{files} pointer, the analyst might encounter difficulties in identifying all the file descriptors opened by the process. A similar amount of inconsistencies are also present in the variable-sized arrays that hold pointers to the \texttt{file} structures that hold the file and socket descriptors open by a process (19\%) and the \texttt{fdtable} structures that connect the \texttt{task\_struct} to the variable-sized arrays (15\%).

In the case of Types 1 and 2 inconsistencies, it is possible that a pointer that in memory refers to a type of structure, in the dump file points instead to a different type of structure. This occurs in our dataset in three dumps: $D_4$, $D_5$, and $D_6$. $D_6$, in particular, contains 279 of these cases, mainly related to the \texttt{vm\_file} pointer in the \texttt{vm\_area\_struct} which reference the \texttt{file} structure associated with a memory-mapped file (167 times), the \texttt{vm\_next} pointer (30 times), or the \texttt{rb\_left} and \texttt{rb\_right} pointers of the substructure \texttt{vm\_rb} (45 times). These two last cases are closely related to each other from the point of view of a forensics analyst: \texttt{vm\_next} pointers maintain the linked list of virtual memory areas of a process and are used by various Volatility 2 plugins to explore process memory. Suppose an analyst suspects a \texttt{vm\_next} pointer is corrupted or tampered. In that case, she can use a different Volatility plugin, \texttt{linux\_proc\_maps\_rb}, to explore the virtual memory areas of a process: this plugin walks the red-black tree which links together, along with the linked list, all the \texttt{vm\_area\_struct}. However, our experiments show that 18 \texttt{vm\_area\_struct} in $D_6$ present value inconsistencies in the linked list \textit{and} in at least one field of the red-black tree, while in 2 cases in both fields. In this scenario, the analyst cannot recover the areas of the virtual address space in any way, as all paths used by Volatility 2 are corrupted leading to several issues. For example, the \texttt{linux\_library\_list} module, designed to report all libraries loaded by a process, relies on exploring the \texttt{vm\_area\_struct} linked list to identify loaded libraries. If this linked list is corrupted, as well as the red and black tree, it is possible that a library injected by a malicious process into another might not be identified. For reference, Table~\ref{tab:fields} summarizes the forensically relevant fields that show at least one inconsistency among the top 5 structures ranked by the percentage of affected instances.

Finally, Table~\ref{tab:plugins} in Appendix reports for each plugin considered in our study, the number of tracked fields that contain inconsistencies in our dumps. These affect the output of Volatility 2 plugins in terms of the capability to extract information (causal inconsistencies) and data reliability (value inconsistencies). As the table shows, all plugins are affected by at least one inconsistency with the only exception of \texttt{linux\_psscan}, which, in fact, relies on carving techniques to extract the \texttt{task\_structs} from a dump. As a result, plugins can report incorrect information for all dumps except $D_0$ and $D_1$. This is due to the fact that almost all the essential Volatility 2 plugins need to traverse one of the top 4 data structures per number of inconsistencies. In these cases, as suggested by Pagani et al. in~\cite{pagani2019back}, it would be beneficial to use alternative paths to traverse a `more stable' kernel data structure. By relying on multiple paths to reach the same information, the analyst might be able to identify and overcome Type 1 and 2 inconsistencies. However, Type 3 and 4 inconsistencies would still not be detectable automatically since paths through these types of inconsistencies would produce wrong but plausible results that would be hard to identify in an automated fashion.

\vspace{0.2cm}
\takeaways{Our analysis demonstrates that inconsistencies in the data
structures used in memory forensics are pervasive.
In particular, four fundamental data
structures critical for forensic analysis 
(\texttt{task\_struct}, \texttt{vm\_area\_struct}, \texttt{dentry} and
\texttt{file}) consistently exhibited inconsistencies across all memory
dumps we analyzed. These data structures are used by the majority of
Volatility plugins, and our findings indicate that inconsistencies, in
particular in the \texttt{task\_struct}, make the majority of them either 
unreliable or unusable. \\
We also observed cases where semantically
related pointers within the same structure simultaneously exhibited
anomalies, further complicating the use of plugins that rely on alternative
paths or pointer cross-validation to overcome possibly inconsistencies.
As a result, the only type of Volatility plugins that proved to be more resilient to
inconsistencies are those that use carving
techniques to extract data structures from memory without relying
on pointers to locate them.}

\section{Related Works}
\label{sec:related_works}
Over time, numerous studies~\cite{kornblum2007using, libster2008proposal, hay2009live, moser2013hunting} have pointed out the challenges posed by the lack of atomicity in memory dumps. \textit{Vomel and Freiling}~\cite{vomel2012correctness} were the first 2010 to introduce the concepts of atomicity, integrity, and correctness of a dump. Later, \textit{Gruhn and Freiling}~\cite{gruhn2016evaluating} evaluated acquisition tools based on these criteria, while in 2021, \textit{Freiling et al.} extended the concept to disk snapshots\cite{freiling2021defining}. At the same time, \textit{Case and Richard}, instead, highlighted the pressing issue of page smearing~\cite{case2017memory}. In 2019, \textit{Pagani et al.}~\cite{pagani2019introducing} introduced the "temporal dimension" in memory forensics to give the analyst an initial means of the atomicity of the data structures in the dump. They also show that page smearing introduces inconsistencies in page tables and impacts any analysis involving user space data structures. Furthermore, \textit{Sudhakaran et al.}~\cite{sudhakaran2022evaluating} have also studied the problem in user-space Android apps. Recently, instead, \textit{Ottmann et al.}~\cite{10.1145/3628600} and \textit{Rzepka et al.}~\cite{10.1145/3680293} have adopted a technique based on a custom "pivot process" that tracks internal memory changes with vector clocks, letting the authors detect whether the memory dump preserves causal consistency.

Among the first to propose a solution to the problem of non-atomicity of memory dumps were \textit{Huebner et al.}~\cite{huebner2007persistent} that suggest a kernel redesign to implement an automatic periodic acquisition of the state of the kernel and user applications by the operating system itself. Another solution was developed by \textit{Schatza and Bradley}~\cite{schatz2007bodysnatcher} involving the injection of a minimal kernel that stops the execution of the running OS and dumps its memory. This approach, however, requires a tight and preexistent integration between the dumping kernel and the original OS. In 2009, an approach based on the data remanence effect in memory chips after the reboot of the system was used by Forenscope~\cite{halderman2009lest} to acquire the memory in an atomic way. This technique, known as cold boot memory acquisition, has been proven, unfortunately, to fail on specific chipset and memory banks setups~\cite{carbone2011depth}.

The following year \textit{Martignoni et al.} have introduced HyperSleuth~\cite{martignoni2010live}, a custom hypervisor injected at runtime to perform atomic memory dumps using dump-on-write and dump-on-idle. This promising technology, however, is not applicable in cases where another hypervisor is already running, such as in Windows 10 virtual secure mode, or if the CPU virtualization extensions are not enabled in the BIOS at boot time. Other solutions developed to obtain atomic dumps involve using modified firmware to run the dump tool at higher privilege levels~\cite{reina2012hardware} or with the operating system no longer running but with the memory contents still available~\cite{latzo2021bringing}. These two methods, however, require hardware access to the machine to flash system firmware residing on the motherboard. Recently, in \textit{"Katana: Robust, Automated, Binary-Only Forensic Analysis of Linux Memory Snapshots"}~\cite{franzen2022katana}, the authors have collaterally introduced a Linux kernel module that allows atomic memory dumps. However, the module, by temporarily blocking system interrupts, can cause the crash of the kernel modules responsible for hardware management at the end of the dump process.

\section{Impact and Discussion}
With the exception of few research prototypes running in higher-privilege CPU modes (e.g., from SMM~\cite{reina2012hardware}),
all hardware-based and software-based memory acquisition techniques
do not stop the kernel during the acquisition process on a bare-metal machine.

The fact that non-atomic memory acquisitions can lead to inconsistent or corrupted data has been recognized since at least 2005, when the initial prototypes of memory forensics tools were proposed~\cite{smearing}. However, prior reports on 'unusable' memory images have been largely anecdotal, without data to understand the frequency of these errors in practice. For example, although previous research noted errors in the page tables in roughly 20\% of acquired memory images~\cite{case2017memory}, we demonstrate that the issue is much more widespread, with \textbf{every single} memory dump we acquired containing at least one, and often many, such errors.

Our results show that even memory acquired from an idle operating system can contain inconsistent data. While most of these inconsistencies may be irrelevant for forensic investigations, some are significant. Every dump we acquired contain several processes with errors in their page tables, sometimes leading to fragments of one process's memory being mistakenly attributed to another. Even if we limit our analysis to the data structures used by Volatility 2 to extract information from a memory dump, we observed \textit{tens of thousands} of inconsistencies. For certain structures, nearly one in four instances contained errors in fields used by Volatility. Furthermore, our experiments reveal for the first time that the choice of target filesystem and dump technique can significantly impact both the acquisition time and the extent of kernel-modified data structures. 

While these errors may not compromise every single investigation, the fragility of forensic analysis based on such data is concerning.
Even more if we remember that all our results are conservative and represent only a lower bound of the inconsistencies that could be encountered in real-world investigations. 
In fact, while PANDA does not support emulating a system with more than one CPU core,
on modern multi-core systems the number of inconsistencies will likely increase due to the 
higher number of writes per second performed by parallel kernel threads and the 
synchronization mechanisms (e.g., semaphores and locks) required to manage multiple CPU units.

Moreover, in our experiments, we considered the best-case scenario
where the memory
was 25\% occupied, and no user-triggered network activity occurred during
the acquisition process or in the 30 minutes prior. In real scenarios,
memory usage may be higher, leading to more fragmentation and a larger
number of page tables. Additionally, if the system must remain
operational during memory acquisition -- for instance when memory is
acquired from a server as part of threat hunting or incident response
investigations -- this would further increase the number of inconsistencies. 

Finally, even the
size of the memory affects the outcome: larger systems, even with unused
RAM, may store their physical pages more widely apart, thus resulting in greater
temporal distance between their acquisition and therefore again in a larger number of
inconsistencies. Because of all these factors, we expect our results to represent
the best-case scenario, with the number of errors to increase considerably in
real-world settings. More concerning is that, in most cases, the analyst
cannot ascertain the accuracy of the acquired data.

Previous efforts to mitigate this problem offer valuable insights. 
For instance, \textit{Pagani et al.}~\cite{pagani2019introducing} developed a memory dump tool designed to sequentially capture causally-related data structures, rather than indiscriminately acquiring physical pages. 
Although this tool is no longer supported and incompatible with modern Linux kernels, it represents a promising direction for research that could significantly reduce inconsistencies. 
By quantifying the problem, our work provides researchers with a deeper understanding and encourages further exploration into developing more robust solutions for atomic memory acquisition.

\section{Code Availability}
The code of Strata PANDA plugins and datasets are available as an open-source project~\cite{strata}.

\section{Acknowledgments}
This work has benefited from a government grant managed by the National Research Agency under France 2030 with reference
``ANR-22-PECY-0009''.

\bibliographystyle{IEEEtran}
\bibliography{IEEEabrv,biblio}

\appendix

\begin{table}[h!]
  \scriptsize
  \caption{\label{tab:plugins} Fields involved in inconsistencies per plugin.}
\begin{tabular}{lcc}
  \toprule
  \textbf{Plugin} & \textbf{\thead{Fields causal\\ inconsistencies}} & \textbf{\thead{Fields value\\inconsistencies}} \\
  \midrule
  \texttt{linux\_check\_creds} & 1 & -\\
  \texttt{linux\_check\_inline\_kernel} & 9 & 3\\
  \texttt{linux\_check\_syscalls} & 2 & 3\\
  \texttt{linux\_dump\_map} & 1 & 2\\
  \texttt{linux\_elfs} & 9 & 1\\
  \texttt{linux\_enumerate\_file} & 2 & 3\\
  \texttt{linux\_find\_file} & 2 & 3\\
  \texttt{linux\_getcwd} & 5 & -\\
  \texttt{linux\_info\_regs} & 1 & 1\\
  \texttt{linux\_ldrmodules} & 3 & 2\\
  \texttt{linux\_library\_list} & 3 & 1\\
  \texttt{linux\_librarydump} & 3 & 2\\
  \texttt{linux\_list\_raw} & 6 & -\\
  \texttt{linux\_lsof} & 8 & 1\\
  \texttt{linux\_malfind} & 3 & 3\\
  \texttt{linux\_memmap} & 1 & -\\
  \texttt{linux\_mount} & 2 & -\\
  \texttt{linux\_netstat} & 2 & 1\\
  \texttt{linux\_plthook} & 3 & -\\
  \texttt{linux\_proc\_maps} & 12 & 6\\
  \texttt{linux\_proc\_maps\_rb} & 2 & 1\\
  \texttt{linux\_procdump} & 1 & 1\\
  \texttt{linux\_process\_hollow} & 3 & 2\\
  \texttt{linux\_process\_info} & 9 & 4\\
  \texttt{linux\_psaux} & 1 & 1\\
  \texttt{linux\_psenv} & 1 & -\\
  \texttt{linux\_pslist} & 1 & 2\\
  \texttt{linux\_psscan} & - & -\\
  \texttt{linux\_pstree} & 2 & 1\\
  \texttt{linux\_recover\_fs} & 2 & 5\\
  \texttt{linux\_threads} & 3 & 3\\
  \texttt{linux\_tmpfs} & 2 & 1\\
  \texttt{linux\_truecrypt} & 3 & 3\\
  \bottomrule
  \end{tabular}
\end{table}

\end{document}